\documentclass[12pt,english,aps,manuscript]{article}
\usepackage[T1]{fontenc}
\usepackage[latin1]{inputenc}
\usepackage{amsmath}
\usepackage{amssymb}
\usepackage[numbers]{natbib}

\makeatletter

\usepackage{geometry}

\makeatletter

\usepackage{geometry}

\makeatletter

\usepackage{color}

\makeatletter
\newcommand{\bee}{\begin{equation}}
\newcommand{\ee}{\end{equation}}
\newcommand{\beea}{\begin{eqnarray}}
\newcommand{\eea}{\end{eqnarray}}

\makeatother

\makeatother

\makeatother

\usepackage{babel}
\makeatother

\begin{document}
\begin{center}
\textbf{\Large Gauge Mediated Supersymmetry Breaking and String Theory}{\Large {}
} 
\par\end{center}

\begin{center}
\vspace{0.3cm}

\par\end{center}

\begin{center}
{\large S. P. de Alwis$^{\dagger}$ and Z. Lalak$^{*}$} 
\par\end{center}

\begin{center}
\vspace{0.3cm}

\par\end{center}

\begin{center}
\textbf{Abstract} 
\par\end{center}

\begin{center}
\vspace{0.3cm}

\par\end{center}

We discuss the possibility of finding scenarios, within type IIB string
theory compactified on Calabi-Yau orientifolds with fluxes, for realizing
gauge mediated supersymmetry breaking. We find that while in principle
such scenarios are not ruled out, in practice it is hard to get acceptable
constructions, since typically, supersymmetry breaking cannot be separated
from the stabilization of the light modulus.

\vfill{}

$^{\dagger}$ Physics Department, University of Colorado, Boulder,
CO 80309 USA. dealwiss@colorado.edu

{*} Institute of Theoretical Physics, Faculty of Physics, University
of Warsaw, ul. Hoza 69, 00-681 Warsaw, Poland. zygmunt.lalak@fuw.edu.pl

\eject

\section{Introduction}

Gauge mediated supersymmetry breaking (GMSB) (for a review see \citep{Giudice:1998bp})
is usually discussed in the context of global supersymmetry (SUSY)
breaking. However when global SUSY is (spontaneously) broken a cosmological
constant (CC) is generated at the same order as the supersymmetry
breaking. It is only within the supergravity (SUGRA) context that
one has the possibility of fine tuning the CC to zero. Thus any discussion
of GMSB must necessarily be embedded within a supergravity context.
It is then natural to ask whether such a SUGRA embedding can be justified/realized
from string theory. In particular we would like to ask whether the
small parameters in GMSB models (in effect mass scales which are highly
suppressed relative to the Planck scale) can find a natural explanation
in string theory. Within the global context such small mass scales
have been justified by dynamical generation of a small scale in an
asymptotically free gauge theory (in the so-called retrofitted models
\citep{Dine:2006gm}). However this leaves the question of what fixes
the relevant gauge coupling at the high scale, open. In string theory
the gauge coupling (just like all other parameters) is a modulus which
needs to be fixed by the dynamics of the model. Also typically the
high scale in these models is necessarily well below the Planck scale.
This too needs to be identified (in the string theory context) as
either the mass of the lightest Kaluza-Klein state or the lightest
modulus. In this note we explore these possibilities.

There is a large variety of GMSB models in the literature. What we
will do here is to consider the simplest set up which exhibits all
the essential features of these models. The issues that we are interested
in here will be common to all GMSB models so there is no loss of generality
in focusing on this simple set up. As we said before most discussions
of these models are in the context of global SUSY. The embedding of
such models within SUGRA was fist discussed by Kitano \citep{Kitano:2006wz}
who took the Kähler potential %
\footnote{We put $M_{P}=1$ in this work unless otherwise indicated.%
}, \begin{equation}
K=X\bar{X}-\frac{(X\bar{X})^{2}}{\Lambda^{2}}+f\bar{f}+\tilde{f}\bar{\tilde{f}}+K_{MSSM},\label{eq:Kkit}\end{equation}
 and the superpotential\begin{equation}
W=c+\mu^{2}X+\lambda Xf\tilde{f}+W_{MSSM}.\label{eq:Wkit}\end{equation}
 The non-renormalizable second term in $K$ arises from integrating
out states at some high scale $\Lambda$. $X$ is a gauge neutral
chiral superfield which is responsible for SUSY breaking. The superfields
$f,\tilde{f}$ are to be identified as the messengers of GMSB and
are charged under the gauge group. The model has a true minimum (i.e.
with no tachyons or flat directions) with the fields taking values
$X_{0}=\frac{\sqrt{3}\Lambda^{2}}{6},\, f=\tilde{f}=0$ , provided
that $\Lambda^{4}>\frac{12\mu^{2}}{\lambda}$ and the CC is tuned
to zero i.e.\begin{equation}
\mu^{2}\simeq\sqrt{3}c=\sqrt{3}m_{3/2}.\label{eq:cckit}\end{equation}
 Using the standard mass formula for scalar masses in SUGRA the mass
matrices may be evaluated. The scalar messengers have squared masses
\begin{equation}
\frac{\lambda^{2}\Lambda^{4}}{12}\pm\lambda\mu^{2},\label{eq:messmass}\end{equation}
 while the scalar partner of the Goldstino (sGoldstino) has a mass
$m_{X}\simeq2\mu^{2}/\Lambda$ \citep{deAlwis:2010sw}. $ $ Finally,
the SUSY breaking is characterized by \begin{equation}
F^{X}\simeq\mu^{2}=\sqrt{3}m_{3/2},\label{eq:FX}\end{equation}
 so that the relevant mass parameter determining soft terms in GMSB
is (restoring $M_{P}$ for clarity) \begin{equation}
m\sim\frac{\alpha}{4\pi}\frac{F^{X}}{X}\simeq\frac{\alpha}{4\pi}\frac{M_{P}^{2}}{\Lambda^{2}}6\, m_{3/2}.\label{eq:softGMSB}\end{equation}

This simple model illustrates the general requirements of GMSB. Note
that gravity mediated SUSY breaking is always present and gives $m\sim m_{3/2}$.
So in order to have GMSB dominance we need to enhance its contribution
relative to the the gravity mediated one i.e. \[
\frac{\alpha}{4\pi}\frac{M_{p}^{2}}{\Lambda^{2}}\gg1.\]
 However we need to take $\Lambda$ to be even smaller if the corresponding
gravity mediated model has FCNC. Thus we probably need the RHS of
the above inequality to be a factor of $10^{2}$ or so. Thus we need
(assuming $\alpha/4\pi\sim10^{-2}$), $\Lambda<10^{-2}M_{P}$. In
fact, additional phenomenological constraints (which are usually imposed
in the GMSB literature) suggest that we need to take a value which
is much smaller.

Let us discuss these phenomenological constraints. We need $m\sim100{\rm GeV}-1{\rm TeV}$,
in order to `solve' the hierarchy problem. The above considerations
would then suggest that we could have a value of $m_{3/2}\sim1{\rm GeV}$.
Such light gravitini were recognized long ago as a source of problems
in standard cosmology, see \citet{Drees:2004jm} for a review. However,
it has been recognized by now that a wide range of gravitino masses
between a few eV and 1-10 GeV can be tolerated depending on the mass
of the NLSP, on the reheating temperature and on nonstandard events
in the history of the universe, see for instance \citet{Feng:2010ij},\citet{Bolz:2000fu},\citet{Covi:2010au}.
In the rest of the paper we will use $m_{3/2}\lesssim1{\rm keV}$
as a ball park figure. This implies \begin{equation}
\Lambda\lesssim10^{-5}M_{P}.\label{eq:Lambdainequlity}\end{equation}
 For comparison we will also include (where relevant) the estimates
for two other values for the gravitino mass: $m_{3/2}\sim1GeV$, corresponding
to the case $\Lambda\sim10^{-2}M_{P}$ mentioned above, and $m_{3/2}\sim1eV$
(leading to $\Lambda\sim10^{-7}M_{P}$). As it turns out, the main
conclusion doesn't depend on a particular value of $m_{3/2}$, only
numerical details do change. 

In more general models of the hidden sector the situation can be even
more severe. As observed in \citep{Lalak:2008bc}, \begin{equation}
m_{soft}=\frac{\alpha}{4\pi}\frac{F^{X}}{\Lambda^{\gamma}},\end{equation}
 where $\gamma=2$ for the Kitano model (order four correction to
the Kähler potential balanced by a gravitational correction), $\gamma=1$
if R-symmetry becomes spontaneously broken due to radiative corrections
in a globally supersymmetric model and $\gamma=4/3$ if the R-symmetry
breaking is caused by an order six correction to a Kähler potential
balanced by a gravitational correction. Assuming a $1$ keV gravitino,
$\gamma=1$ results in $\Lambda\lesssim10^{-10}M_{P}$ and $\gamma=4/3$
in $\Lambda\lesssim10^{-7.5}M_{P}$. For the sake of concretness we
are going to assume the conservative value $\gamma=2$.

The question we will address in the rest of the paper is, can we integrate
out in a supersymmetric way all string theory moduli, to produce an
effective theory with the scale of the lightest particle that is integrated
out having a mass $\Lambda$ which is several orders of magnitude
below the Planck scale. The essential point here is that the supersymmetry
breaking mechanisms postulated in connnection with GMSB, are essentially
four dimensional effective theories, which in a string theory context,
are effectively obtained after decoupling all the string theory moduli,
which therefore need to have been stabilized with their masses at
some high scale. This is the only viable interpretation of the low
energy supersymmetry breaking models that have been considered in
the context of GMSB, when the question of their relation to string
theory is considered. Otherwise one would have to look at the full
effective theory coming from string theory, treating the moduli along
with the fields of the low energy theory as dynamical variables and
this would certainly vitiate the arguments made in the context of
the low energy supersymmetry breaking models. Thus the necessary criterion
for embedding these models within string theory, is that the highest
scale of the effective theory (i.e. $\Lambda$) should be lower than
the mass of the lightest string theory modulus. We stress here that
this requirement does not come from a cosmological (or phenomenological)
constraint but from the need to construct an effective low energy
theory from which all the moduli of string theory have been integrated
out.

\section{KKLT model}

The simplest possibility is to integrate out the moduli/dilaton in
type IIB in a KKLT \citep{Kachru:2003aw} like situation i.e. with
a SUSY AdS minimum, and then tune the model of the previous section
to uplift the CC. In particular this means that the gravitino mass
(CC) at the KKLT stage should be comparable to the final gravitino
mass i.e. $m_{3/2}\sim10^{-24}$.

\begin{eqnarray}
K & = & -3\ln(T+\bar{T})-k(U,\bar{U})-\ln(S+\bar{S}),\label{eq:K0}\\
W & = & A(U)+SB(U)+Ce^{-aT},\label{eq:WKKLT}\end{eqnarray}
 with $U=\{U^{a},\, a=1,\ldots,h_{12}\}$. The first two terms in
$W$ are generated by internal fluxes while the third is a non-perturbative
(NP) contribution. The KKLT minimum preserves SUSY - all the moduli
are determined by the equations\begin{equation}
D_{S}W=D_{a}W=D_{T}W=0.\label{eq:KKLT}\end{equation}
 Denoting by $W_{flux}$ the value of the first two terms in \eqref{eq:WKKLT}
we have, from the last equation above

\[
aCe^{-aT}=\frac{W_{flux}}{T+\bar{T}},\]
 and\[
m_{3/2}\sim\frac{W_{0}}{(T+\bar{T})^{3/2}}\sim(T+\bar{T})^{-1/2}aCe^{-aT}.\]
 In these equations $T$ is the value of this modulus at the minimum.

As we pointed out before, in order to avoid additional fine tuning,
we must take this to be at least as small as the final gravitino mass
after uplift and SUSY breaking from the Polonyi type models of the
previous section - i.e. we need the LHS of the above relation to be
$\sim10^{-24}$.

On the other hand the mass of the Kähler modulus is\[
m_{T}\sim\sqrt{T}aCe^{-aT},\]
 so that \begin{equation}
\frac{m_{T}}{m_{3/2}}\sim T.\label{eq:mTm3half}\end{equation}
 The minimal requirement for getting Kitano models from this string
construction is to have $m_{T}\gtrsim\Lambda\sim10^{-5}$. Given the
phenomenological bound $m_{3/2}$ $<10^{-24}$, this gives a value
$T\gtrsim10^{19}$. However this would give a string scale (note that
the volume of the CY is $\sim1/T^{3/2}$) $m_{string}\sim M_{P}/\sqrt{T^{3/2}}\sim10^{-14}M_{P}$.
But this is much less than $\Lambda$ so that the model is inconsistent.
In general the condition $m_{T}\gtrsim\Lambda$ leads to \begin{equation}
T>10^{19}\left(\frac{m}{100\,{\rm GeV}}\,\frac{m_{3/2}}{1\,{\rm keV}}\right)^{-1/2},\end{equation}
 hence raising $m$ and $m_{3/2}$ helps only slightly.  Note that
even with $m_{3/2}\sim1GeV=10^{-18}M_{P}$ we would get $m_{T}\gtrsim\Lambda\sim10^{-2}$
(see para after \eqref{eq:Lambdainequlity}) which will give $T\sim10^{16}$
and hence $m_{string}\sim10^{-12}\ll\Lambda$. With a very light gravitino
on other hand ($m_{3/2}\sim1eV$, $\Lambda\sim10^{-7}M_{P}$) the
above calculation gives $T\sim10^{20}$ and $m_{string}\sim10^{-15}M_{P}\ll\Lambda$.
Clearly the model is inconsistent for any acceptable value for $m_{3/2}$
in GMSB.

\section{KKLT racetrack model in IIB}

The situation is considerably improved when we consider racetrack
models within the KKLT context. The model is defined by

\begin{eqnarray}
K & = & -3\ln(T+\bar{T})-k(U,\bar{U})-\ln(S+\bar{S}),\label{eq:K1}\\
W & = & A(U)+SB(U)+\sum C_{i}e^{-a_{i}T},\label{eq:W1}\end{eqnarray}
 with $U=\{U^{a},\, a=1,\ldots,h_{12}\}$. Assume $C_{i}$ are constants
i.e. ignore threshold effects. Obviously we need at least two NP terms.
For condensing $SU(N)$ groups $a_{i}=2\pi/N_{i}$.

Let us look for SUSY Minkowski minima: $\partial W=W=0$,\begin{eqnarray}
\partial_{S}W & = & B(U)=0,\,\partial_{T}W=-\sum_{i}a_{i}C_{i}e^{-a_{i}T}=0,\label{eq:dSdT=0}\\
\partial_{a}W & = & A_{a}(U)+B_{a}(U)S=0.\label{eq:dU=0}\end{eqnarray}
 First and last equations are $h_{21}+1$ relations determining $U^{a}=U_{0}^{a},\, S=S_{0}$.
In the simplest case where we have just two NP terms the second equation
gives\begin{equation}
e^{2\pi\frac{N_{2}-N_{1}}{N_{1}N_{2}}T_{0}}=-\frac{N_{2}}{N_{1}}\frac{C_{1}}{C_{2}}.\label{eq:expT}\end{equation}
 With any solution of these equations for the moduli/dilaton, fluxes
may be chosen such that $W(T_{0},U_{0},S_{0})=0$.

Given a large number of 3-cycles, and the large number of choices
of fluxes through them, one expects that the fine tuning of $W$ to
zero is compatible with having the masses of the $U^{a},$ moduli
and of axio-dilaton $S$ being very large - close to the string scale
say. The lightest modulus would as usual be $T$ since it gets its
mass from the NP terms, \begin{equation}
m_{T}^{2}K_{T\bar{T}}|_{0}=e^{K}|\partial_{T}^{2}W|^{2}K^{T\bar{T}}|_{0}M_{P}^{2}\sim\frac{1}{T_{0}+\bar{T}_{0}}|M_{P}^{2}\sum_{i}(\frac{2\pi}{N})^{2}e^{-2\pi T_{0}/N}|^{2}.\label{eq:mT2KTT}\end{equation}
 In $ $the last relation we have estimated the size of the Kähler
potential terms from $U,S$ to be of $O(1)$. Note that $K_{T\bar{T}}=3/(T+\bar{T})^{2}$.
The following results are simple consequences of this set up:

\begin{itemize}
\item The KK scale is $M_{KK}\sim M_{P}/\Re T_{0}$ and the consistency
of the whole framework requires that this mass be greater than $m_{T}$.
This implies that \begin{equation}
m_{T}\sim(T_{0}+\bar{T_{0}})^{1/2}|\sum_{i}\left(\frac{2\pi}{N_{i}}\right)^{2}e^{-2\pi T_{0}/N_{i}}|M_{P}<\frac{M_{P}}{(T_{0}+\bar{T}_{0})}.\label{eq:mT}\end{equation}
 Now the phenomenology of GMSB requires \eqref{eq:Lambdainequlity}.
If we identify $\Lambda$ with the KK mass this requires $T_{0}\gtrsim10^{5}$. 
\item If we identify $\Lambda$ with the lower mass $m_{T}$ clearly we
need a value of $T$ which is parametrically larger than the Planck
scale. This implies that we need large values of $N_{1},N_{2}$ so
that from \eqref{eq:expT} we have (taking $N_{2}>N_{1})$ \begin{equation}
\tau_{0}\sim\frac{N_{2}}{2\pi},\label{eq:T0}\end{equation}
 where $\tau_{0}$ is the real part of $T$ at the minimum. This gives
\begin{equation}
m_{T}\sim(2\pi/N_{2})^{3/2}M_{P}\sim M_{P}/\tau_{0}^{3/2}.\label{eq:mT2}\end{equation}
 Identifying this with $\Lambda\lesssim10^{-5}M_{P}$ gives, $\tau_{0}\gtrsim3\times10^{3}$.
This however requires very large gauge groups\begin{equation}
N_{1},N_{2}\gtrsim10^{4}.\label{eq:N1N2}\end{equation}

\end{itemize}
These are rather large values, but in certain F-theory constructions
\citep{Candelas:1997eh} such large groups have in fact been obtained,
though it is not clear to us how common they are in the landscape
of IIB orientifold compactifications (or the corresponding F-theory
constructions).

Note that the estimate in \eqref{eq:T0} and hence \eqref{eq:N1N2}
were made under the assumption that any difference in the ratio of
the pre-factors $C_{1}/C_{2}$ from unity should disappear for large
ranks, i.e. $|\frac{C_{1}}{C_{2}}|-1\lesssim O(1/N)$ since the only
difference between them comes from the $O(1)$ difference between
the ranks. The explicit dependence on the rank (which is governed
by the number of D7 branes wrapping the four cycle) calculated in
\citep{Baumann:2006th} certainly satisfies this. If however for some
reason there are large threshold effects which make $C_{1}/C_{2}$
significantly larger than $1$ then these estimates would be changed.
In this case we would have $\tau_{0}\sim\frac{N^{2}}{2\pi},$ so that
$N_{1},N_{2}\sim O(10)$. It is an interesting problem to propose
a source of such large threshold corrections in the string theoretic
context.

\section{LVS models}

A natural way of getting a large volume (which seems to be needed
to explain $\Lambda\lesssim10^{-5}M_{P}$) is to consider LVS models
\citep{Balasubramanian:2005zx}. These need at least two Kähler moduli
with one of them being a blow-up modulus. The simplest example has
the CY volume being of the form \[
{\cal V}=\tau_{b}^{3/2}-\tau_{s}^{3/2}.\]
 Here $\tau_{b,s}$ govern the size of the large 4-cycle and the blown-up
four cycle respectively. When the potential in minimized we get \begin{eqnarray}
e^{-a\tau_{s}} & \simeq & \frac{3}{4}\frac{W_{0}}{aA{\cal V}}\sqrt{\tau_{s}}\left(1-\frac{3}{4a\tau_{s}}\right),\label{eq:sol1}\\
\tau_{s}^{3/2} & \simeq & \frac{\hat{\xi}}{2}(1+\frac{1}{2a\tau_{s}}).\label{eq:sol2}\end{eqnarray}
 Here\begin{equation}
\hat{\xi}=-\frac{1}{g_{string}^{3/2}}\chi\zeta(3)/2(2\pi)^{3}\label{eq:xihat}\end{equation}
 where $\chi$ is the Euler character of the manifold and $\zeta$
is the Riemann zeta function. Noting that $m_{3/2}\sim W/{\cal V}$
the first equation gives\begin{equation}
a\tau_{s}\sim|\ln m_{3/2}|\label{eq:taugrav}\end{equation}
 The LVS solution breaks SUSY, hence in order to be compatible with
GMSB we have to tune the gravitino mass to be at or below the GMSB
favored value $m_{3/2}\lesssim1KeV=10^{-24}M_{P}$. So from \eqref{eq:taugrav}
we need $a\tau_{s}\sim10^{2}$ which implies (given that $a\sim O(2\pi)$
at most) from \eqref{eq:sol2} that $\hat{\xi}\sim O(10)$ and hence
from \eqref{eq:xihat} we need CY orientifolds of large Euler character
($\chi\sim5000$). While there are no known barriers to having such
large Euler characters, the largest found so far have values that
are around an order of magnitude smaller. Of course given the logarithmic
dependence on $m_{3/2}$ the above conclusion will remain true for
the whole range of values for this mass discussed in the introduction.

In any case in this type of scenario the light modulus has a mass
$m\sim m_{3/2}/\sqrt{{\cal V}}$ which is obviously below the messenger
scale and hence cannot be integrated out to give the type of action
(like the Kitano model) that GMSB needs.

\section{Retrofitted models}

The model of supersymmetry breaking in \eqref{eq:Kkit}\eqref{eq:Wkit}
has three parameters that are highly suppressed relative to the Planck
scale. In the previous section we identified $\Lambda\sim10^{-5}$
as the scale set by the lowest modulus mass which is integrated out.
The other two parameters $c,\mu^{2}$ have to be fine tuned against
each other to get zero CC after SUSY breaking, and they have to be
extremely tiny in order to get a $1\,{\rm keV}$ (or even $1\,{\rm GeV}$)
gravitino mass. This means that we need \begin{equation}
\mu^{2}\simeq\sqrt{3}c=\sqrt{3}m_{3/2}\sim O(10^{-24}).\label{eq:muctuning}\end{equation}
 How does such a small parameter arise. In retrofitted models \citep{Dine:2006gm}
this is related to a dynamically generated scale as in QCD. In the
context of SUSY QCD this may be identified with gaugino condensation.
So we effectively have\begin{equation}
\mu^{2}=\Sigma^{3}=\Lambda^{3}\exp\left(-\frac{3}{b}\frac{8\pi^{2}}{g^{2}(\Lambda)}\right).\label{eq:Sigma}\end{equation}
 Here $\Sigma$ is the dynamically generated scale, $\Lambda$ is
the UV scale of the effective theory that we identified in the previous
section, $g(\Lambda)$ is the coupling at that scale of some gauge
group, and $b$ is the corresponding one-loop beta-function coefficient
(for example $b=3N$ for pure $SU(N)$ Yang-Mills). Even for very
small gauge groups with say $N=2$ this requires a gauge coupling
constant at the UV scale to be quite small $\alpha=\frac{g^{2}}{4\pi}\sim0.06$
to satisfy the estimate \eqref{eq:muctuning}.

In string theory however this coupling constant is actually a field.
As we discussed in the previous section it is most naturally identified
with the modulus $T$. Then we might try to identify the supersymmetrty
breaking sector (Polonyi field) with the open string moduli (for simplicity
represented by a single field $X$ here), that describe the location
of the stack of D3 branes relative to the stack of D7 branes which
wrap the four cycle in the CY space, and generate the NP term in \eqref{eq:W1}.
Following Baumann et al \citep{Baumann:2006th} (in particular its
adaptation to our situation in \citep{deAlwis:2007qx}) we should
modify this term to read\begin{equation}
\sum A(1-\frac{X}{\sigma})^{1/N_{i}}e^{-\frac{2\pi}{N_{i}}T},\label{eq:Phi}\end{equation}
 where $\sigma$ represents the location of the D7 stack in the internal
space. The natural scale at which to make the identification the guage
coupling $\alpha=T$, would be the Planck scale. In this case clearly
we cannot produce the small number that is necessary to satisfy \eqref{eq:muctuning}
since the exponential factor is fixed by the racetrack to be $\exp\{-O(1)\}$.
Even if this identification is made at the cut off scale $\Lambda$,
where we would have $\alpha(\Lambda)=T$ and $A=\Lambda^{3}$, the
exponential factor would still need to produce a factor of around
$10^{-9}$ relative to $\Lambda^{3}$ (see \eqref{eq:Sigma}) to get
a gravitino at the keV scale. This is clearly not possible.

The alternative is to couple $X$ to the theory on a stack of D3 branes.
Perhaps this field can be considered as a modulus describing the location
of this stack relative to the standard model stack which is located
at some singularity. Assuming a condensing gauge group on the stack
of D3 branes, and that the corresponding gauge coupling at some high
scale, which we again take to be $\Lambda$, is $S$ the dilaton,
we have $\alpha(\Lambda)=S$. Unlike in the case discussed above we
do not know of a string construction that accomplishes this. Nevertheless
let us ask whether a term such as \begin{equation}
\Lambda^{3}Xe^{-\frac{2\pi}{N}S}\label{eq:SNP}\end{equation}
 can account for the $\mu^{2}X$ term in the Polonyi model. To get
the appropriate suppression (with $\Lambda=10^{-5}$) seems to require
$S=\frac{\partial_{U}A}{\partial_{U}B}\sim5-10$ with low values of
$N$. On the other hand the mass of the dilaton is given by \[
m_{S}\sim\sqrt{s}\frac{|\partial_{U}B|}{\tau_{0}^{3/2}}.\]
 This is to be compared with the mass $m_{T}\sim1/\tau_{0}^{3/2}$
(see \eqref{eq:mT2}). Clearly fluxes can be chosen such that the
masses satisfy the consistency condition $m_{S}>m_{T}\sim\Lambda$.
If on the other hand (as is more natural in SUGRA) we identified the
high scale at which the boundary value of the coupling is equal to
$S$ as the Planck scale (so effectively replacing the pre-factor
$\Lambda^{3}$ by $1$ in \eqref{eq:SNP}) we would need (with low
values of $N$) values of $S$ of $O(10)$. Again there does not appear
to be any problem with this.

\section{Models with charged moduli - role of D-terms}

Finally let us consider a scenario where all but the Kähler modulus
have been integrated out close to the string scale. In this case an
interesting class of models corresponds to the situation, when this
modulus is charged under an anomalous $U(1)$ gauge group (equivalently
one can say that the shift of the modulus has been gauged). Then the
expectation is that the modulus becomes a component of the massive
vector multiplet and, indeed, decouples from the light spectrum, see
\citep{Lalak:1999bk}. This question has been studied for example
in the last reference, and more recently in \citep{Cvetic:2008mh}
(in the context of Type I string theory, with a D1-instanton inducing
the Polonyi term) and in \citep{Jelinski:2009cp}. To be more specific,
let us consider the superpotential \begin{equation}
W=W_{0}+fXe^{-T},\end{equation}
 invariant under an $U(1)$. For the sake of discussion let us first
consider the Kähler potential \begin{equation}
K=|X|^{2}-\frac{|X|^{4}}{\Lambda^{2}}+\frac{m_{V}^{2}}{2}(T+T^{\dagger}-V)^{2},\end{equation}
 where $m_{V}$ is the mass of the gauge boson of the $U(1)$. In
addition to the F-term scalar potential one finds also a non-trivial
D-term contribution of the form \begin{equation}
V_{D}=\frac{1}{2}\left((|X|^{2}-2\frac{|X|^{4}}{\Lambda^{2}})+m_{V}^{2}(T+T^{\dagger})\right)^{2}.\end{equation}
 One finds that the expectation value of $X$ is fixed as in the Kitano
model and the D-term fixes the expectation value of the modulus at
$<T>=\frac{\Lambda^{4}}{24m_{V}^{2}(1+3\frac{\Lambda^{2}}{m_{V}^{2}})^{2}}$
\citep{Jelinski:2009cp}. The mass of the real part of $T$ is $2m_{V}$
while the imaginary part of the modulus is swallowed by the massive
gauge boson. Taking $m_{V}$ of the order of the compactification
scale one concludes that the modulus decouples. However, the resulting
expectation value of the modulus depends crucially on the form of
its Kähler potential. In the above example we have assumed a quadratic
Kähler potential for the T, which implies a small expectation value
and an unsuppressed effective mass scale multiplying $X$ in the Polonyi
term. Unfortunately, for the Kähler modulus the appropriate form is
logarithmic, $K=-3\log(T+T^{\dagger})$. This in turn gives $\sim\frac{3}{T+T^{\dagger}}$
in the D-term, which implies a run-away vacuum solution for $T$,
since the term in $V_{D}$ which depends on $X$ must be very small.
Large values of $<T>$ suppress very strongly the effective mass scale
in the Polonyi term making the model unrealistic (see also \citep{Dudas:2008qf}).

\section{Conclusions}

We have found that (within the well understood context of IIB flux
compactifications), simple models which allow for GMSB like scenario
below the lowest modulus mass scale are possible, but they are rather
unlikely in the context of the landscape. In LVS type compactifications
\citep{Balasubramanian:2005zx} the string theory sector cannot be
integrated out since there is a light modulus that is below the messenger
scale. The simplest KKLT type model (i.e. with just one non-perturbative
exponential term) will not work since when the necessary constraints
for realizing a GMSB scenario are imposed, we get a string scale that
is below the messenger scale and so is inconsistent. The only possibility
(as far as we can see) is to have an extended KKLT scenario, with
more than one NP term, i.e. a racetrack. In this case we can get the
necessary conditions, however the resulting gauge groups are anomalously
large, though it may be possible to obtain them in some F-theory constructions.
Even if that were the case, the simplest possibility for explaining
the low value of the Polonyi scale (i.e. the SUSY breaking scale)
does not work, and while we have suggested an alternative it is not
clear that this can actually be realized in string theory. Finally
we made some comments about the possibility of D-term effects and
concluded that they will not help if the Kähler potential for the
relevant modulus is of the form that one usually gets in string theory.

\section{Acknowledgments}

The research of SdA is partially supported by the United States Department
of Energy under grant DE-FG02-91-ER-40672. The research of ZA is partially
supported by Polish Ministry for Science and Education under grant
N N202 091839. Both of us wish to thank KITP, Santa Barbara for hospitality
while this work was in its formative stages.

\bibliographystyle{apsrev} \bibliographystyle{apsrev}
\bibliography{myrefs}
 
\end{document}